\documentclass[structabstract]{aa}  

\usepackage{natbib}	
\usepackage{amsmath}
\usepackage{amssymb}
\usepackage{txfonts,epsfig,graphicx,url,twoopt, subfigure}
\usepackage[breaklinks=true]{hyperref} 
\hypersetup{colorlinks=true,citecolor=blue}
\bibpunct{(}{)}{;}{a}{}{,}   
\usepackage[usenames,dvipsnames]{color}

\begin{document}

\title{Gas and dust structures in protoplanetary disks hosting multiple planets}
   \author{P.~Pinilla\inst{1}, M.~de~Juan~Ovelar\inst{2}, S.~Ataiee\inst{3}, M.~Benisty\inst{4}, T.~Birnstiel\inst{5}, E.~F.~van~Dishoeck\inst{1,6}, M.~Min \inst{7}}
   	\institute{Leiden Observatory, Leiden University, P.O. Box 9513, 2300RA Leiden, The Netherlands\\
   	\email{pinilla@strw.leidenuniv.nl}
	\and
	Astrophysics Research Institute, Liverpool John Moores University,146 Brownlow Hill, Liverpool L3 5RF, UK
	\and
	School of Astronomy, Institute for Research in Fundamental Sciences,Tehran, Iran
	\and
	Univ. Grenoble Alpes, IPAG, F-38000 Grenoble, France\\ CNRS, IPAG, F-38000 Grenoble, France 
	\and
	Harvard-Smithsonian Center for Astrophysics, 60 Garden Street, Cambridge, MA 02138, USA
	\and
	Max-Planck-Institut fŸr Extraterrestrische Physik, Giessenbachstrasse 1, 85748, Garching, Germany 
	\and
	Astronomical Institute Anton Pannekoek, University of Amsterdam, PO Box 94249, 1090 GE Amsterdam, The Netherlands}
   \date{Received  25 July 2014/ Accepted 17 October 2014}
 \abstract
{Transition disks have dust-depleted inner regions and may represent an intermediate step of an on-going disk dispersal process, where planet formation is probably in progress. Recent millimetre observations of transition disks reveal radially and azimuthally asymmetric structures, where micron- and millimetre-sized dust particles may not spatially coexist. These properties can be the result of  particle trapping and grain growth in pressure bumps originating from the disk interaction with a planetary companion. The multiple features observed in some transition disks, such as SR~21, suggest the presence of more than one planet.}
{We aim to study the gas and dust distributions of a disk hosting two massive planets as a function of different disk and dust parameters.  Observational signatures, such as spectral energy distributions, sub-millimetre, and polarised images, are simulated for various parameters.} 
{Two dimensional hydrodynamical and one dimensional dust evolution numerical simulations are performed for a disk interacting with two massive planets. Adopting the previously determined dust distribution, and assuming an axisymmetric disk model, radiative transfer simulations are used to produce spectral energy distributions and synthetic images  in polarised intensity at 1.6 $\mu$m and sub-millimetre wavelengths (850~$\mu$m). We analyse possible scenarios that can lead to gas azimuthal asymmetries.} 
{We confirm that planets can lead to particle trapping, although for a disk with high viscosity ($\alpha_{\rm{turb}}=10^{-2}$), the planet should be more massive than $5~M_{\rm{Jup}}$ and dust fragmentation should occur with low efficiency ($v_{f}\sim30\rm{m~s}^{-1}$). This will lead to a ring-like feature as observed in transition disks in the millimetre. When trapping occurs, we find that a smooth distribution of micron-sized grains throughout the disk, sometimes observed in scattered light, can only happen if the combination of planet mass and turbulence is such that small grains are not fully filtered out.  A high disk viscosity ($\alpha_{\rm{turb}}=10^{-2}$) ensures a replenishment of the cavity in micron-sized dust, while for lower viscosity ($\alpha_{\rm{turb}}=10^{-3}$), the planet mass is constrained to be less than $5~M_{\rm{Jup}}$. In these cases, the gas distribution is likely to show low-amplitude azimuthal asymmetries caused by disk eccentricity rather than by long-lived vortices.}
{}
 

\authorrunning{P.~Pinilla et al.}

\maketitle

\section{Introduction}     \label{introduction}

Planet formation occurs in the gaseous and dusty protoplanetary disks that remain around young stars during their formation. The wide diversity of exoplanets detected in recent decades \citep[e.g.][]{batalha2013} may have its origin in the different gas and dust distributions observed in protoplanetary disks \citep[e.g.][]{mordasini2012}. The understanding of how protoplanetary disks disperse and form planets is a rapidly evolving subject and many questions remain open \citep[e.g.][]{alexander2013}.  Circumstellar disks with little or no excess emission at the near/mid-infrared, but with significant excess at longer wavelengths were first detected in 1989 \citep{strom1989}, indicating disks with dust depleted inner regions. These inner cavities have more recently been directly imaged in the sub-millimetre regime \citep[e.g.][]{brown2009}. 

These so-called transition disks are believed to be an intermediate step of the evolutionary phase between gas/dust-rich disks and debris disks, however, whether or not transition disks are the result of undergoing disk dispersal is still a subject of discussion \citep[e.g.][]{espaillat2014}.  Different mechanisms have been proposed to explain the observed properties of transition disks such as protoevaporation \citep[e.g.][]{alexander2006, owen2012}, interaction with one or multiple planets \citep[e.g.][]{varniere2006, dodson2011},  and dust growth \citep[][]{dullemond2005, birnstiel2012}. Additional studies have been done assuming that different processes simultaneously take place in these disks \citep[e.g.][]{rosotti2013}, however, a full understanding of the physical properties of transition disks is far from complete.

\begin{figure*}
 \centering
   \includegraphics[width=18.0cm]{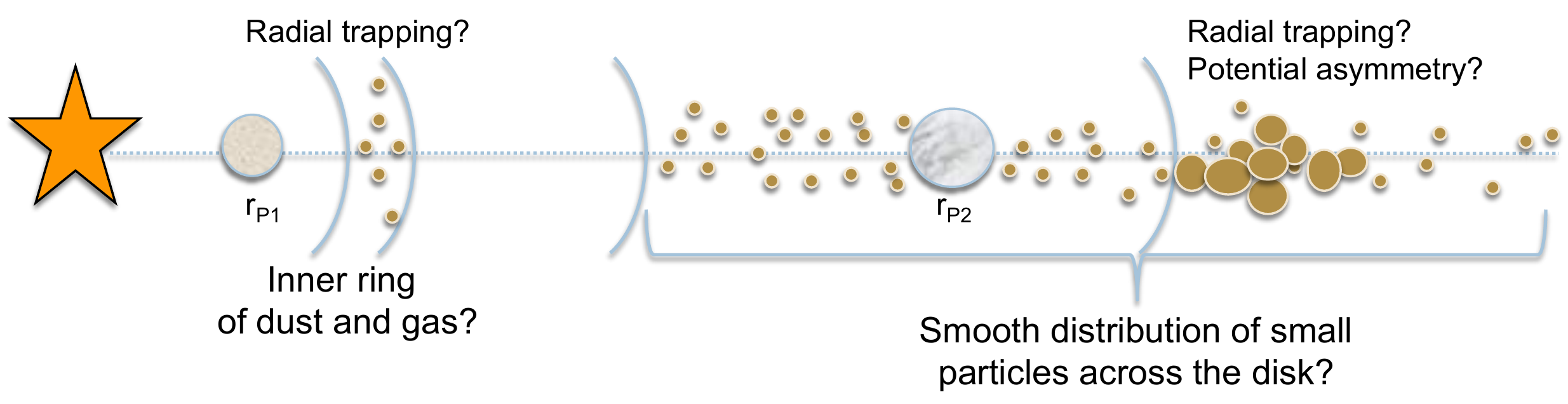}
   \caption{Sketch of the spatial segregation between small and large grains expected when two planets interact with the disk and lead to particle traps. Asymmetric structures may exist due to azimuthal trapping in vortices or/and disk eccentricity.}
   \label{cartoon}
\end{figure*}

Any clearing mechanism that creates a gas surface density depletion leads to a decoupling of small (micron-sized particles) and large grains (millimetre-sized particles) \citep{rice2006, pinilla2012, gonzalez2012, zhu2012}, which may result in dissimilar observed structures at different wavelengths. Observations of transition disks by the Strategic Explorations of Exoplanets and Disks with Subaru survey (SEEDS) show that for some cases, a cavity is not detected at near-infrared scattered light emission, contrary to millimetre observations \citep{dong2012}.  Molecular  gas and polycyclic aromatic hydrocarbon (PAH) emission have also been detected inside millimetre dust cavities \citep{geers2007, salyk2009, pontoppidan2008, maaskant2013b}.  This ``missing cavities'' problem could be the result of considering the evolution of dust when a massive planet is interacting with a disk \citep[e.g.][]{rice2006, dejuanovelar2013}. In these models, the radial distribution of micron-sized particles is expected to be similar to the gas distribution, while a large radial separation between the gas and dust is expected for millimetre particles \citep{pinilla2012}. This interesting radial disentanglement has already been observed for different transition disks, such as HD~135344B  \citep{garufi2013}, HD~100546 \citep{ardila2007, walsh2014}, and SR~21 \citep{follette2013, perez2014}.  Moreover, observations show interesting structures of transition disks, such as  spiral arms at scattered light \citep[e.g.][]{muto2012, grady2013}, eccentric gaps  \citep[e.g.][]{thalmann2014}, and strong azimuthal asymmetries at the millimetre emission \citep{casassus2013, marel2013, fukagawa2013}, which may also result from  planet disk-interaction(s). 

The knowledge of  gas and dust density distributions at different wavelengths may hint at the physical properties of potential embedded  planet(s) in the disk. In this paper, we aim to study the observational signatures in the case of two planets interacting with a disk, by analysing the resulting images and spectral energy distributions (SEDs) obtained by combining hydrodynamical, dust evolution, and radiative transfer numerical simulations. The main questions to address are illustrated in Fig.~\ref{cartoon}:

\begin{itemize}
\item Considering different disk and planet properties, how efficient is dust trapping, and where are the dust traps located?
\item What is the resulting dust size distribution across the disk? Can it lead to spatial segregation between micron- and mm-grains?
\item What are the conditions to induce an asymmetry in the disk? Is it due to vortex formation or disk eccentricity?
\end{itemize}

Thanks to the wealth of available multi-wavelength observations, the disk around SR~21 is an excellent candidate to apply these models. The star SR~21 is located in the Ophiuchus star-forming region, whose disk was identified as a transition disk by \cite{brown2007}.  They reported  Infrared Spectrograph (IRS) data of the Spitzer Space Telescope, and  inferred a 0.45-18~AU gap by modelling the SED. Observations of CO ro-vibrational transition with CRIRES on the Very Large Telescope (VLT)  indicates a narrow ring of gas at $\sim~7$~AU distance from the star \citep{pontoppidan2008}, with a width of $\sim~1$~AU. \cite{follette2013} present scattered light imaging obtained with Subaru HiCIAO camera in $H$~band (1.6~$\mu$m),  finding that large ($\sim$~mm) and small ($\sim~\mu$m) particles are decoupled. They conclude that the polarised intensity radial profile of their observations is decreasing steeply outwards ($r^{-3}$), with no break at the location of the cavity radii observed at millimetre wavelength \citep[$\sim$~36~AU,][]{andrews2011}. The sub-millimetre observations at $350$~GHz (850~$\mu$m) by \cite{brown2009} and \cite{andrews2011} suggest an asymmetric feature, which was confirmed with ALMA observations at 690~GHz (450~$\mu$m) \citep{perez2014} with an emission contrast of around two. To model this asymmetry,  \cite{perez2014} fit a 2D Gaussian at $\sim46$~AU from the central star. In the following, we will consider the stellar parameters of SR~21, and aim to reproduce the main characteristics of these observations. 

The structure of this paper is: in Sect.~\ref{models}, hydrodynamical, dust evolution, and radiative transfer numerical models are described, with the corresponding set-up. The results of the simulations are shown in Sect.~\ref{results}. We present synthetic images and SEDs in Sect.~\ref{observations}. Sections~\ref{discussion} and \ref{summary} are the discussion and  main conclusions of this work.

\section{Models and setups}     \label{models}
In this section, we describe the codes and setups used for the hydrodynamical, dust evolution, and radiative transfer simulations. Our model parameters are chosen to reproduce transition disks with large millimetre cavities and a narrow ring of material inside the cavity. For this,  we assume two non-migrating planets embedded in the disk. 

\subsection{Hydrodynamical simulations} \label{hydro_sims}

We explore the evolution of the gas surface density of a disk hosting two non-migrating planets, by considering the gravitational effect of the planets onto the disk. For these simulations, we used the 2D version of the fast advection hydrodynamical code \texttt{FARGO} \citep{masset2000}. In this code, the length and mass units are scaled by the planet location and the mass of the central star, respectively. In this case, we use \texttt{FARGO} to solve the Navier-Stokes and continuity equations for a  flaring disk ($H\equiv h/r=c_s/v_k$, with $c_s$ the sound speed and $v_k$ the Keplerian velocity) interacting with two planets. The aspect ratio of the disk is increasing with radius as $H=h_0r^{f}$, with a flaring index $f$ of 0.25 and an aspect ratio $h_0$ of 0.05 at the position of the inner planet. 


\begin{table}
\caption{Model parameters}
\centering   
\tabcolsep=0.08cm                      
\begin{tabular}{c|c}       
Parameters  &  Value \\
\hline
\hline 
\texttt{FARGO}&\\
\hline
\hline
$q_{P1}$& $4\times10^{-4}$\\
$q_{P2}$& $[2\times10^{-3}, 4\times10^{-3}, 6\times10^{-3}]$\\
$r_{P1}$ & 1.0\\
$r_{P2}$ & 3.5\\
$r_{\rm in}[r_{P1}]$ & 0.1\\
$r_{\rm out}[r_{P1}]$&24\\
$\Sigma_0 [M_\star/r_{P1}^{2}]$&$1.60\times 10^{-5}$\\
$h_0$ & 0.05\\
$f$ & 0.25\\
$\alpha_{\rm{turb}}$& $[10^{-2}, 10^{-3}]$\\
$n_r\times n_{\phi}$& $512\times587$\\
\hline
\hline
Dust evolution &\\
\hline
\hline
$M_{\star}[M_{\odot}]$& 2.5\\
$R_{\star}[R_{\odot}]$& 3.15\\
$T_{\star, \rm{eff}}[K]$& 5830\\
$M_{\rm{disk}}[M_{\odot}]$& 0.006\\
{\small gas-to-dust ratio}& 100\\
$r_{P1}$[AU] & 5.0\\
$\Sigma_0(5~\rm{AU})$[$\rm{g~cm}^{-2}$]&$14.2$\\
$\rho_s[\rm{g~cm}^{-3}]$&1.2\\
$v_f[\rm{m~s}^{-1}]$& [10, 30]\\
 \hline
 \hline
MCMax&\\
\hline
\hline
$d$[pc]&135\\
$i[^\circ]$ &15\\
\end{tabular}    
\label{table_inputs}
\end{table}

\begin{figure*}
 \centering
 \tabcolsep=0.1cm 
   \begin{tabular}{cc}   
   	\includegraphics[width=12.8cm]{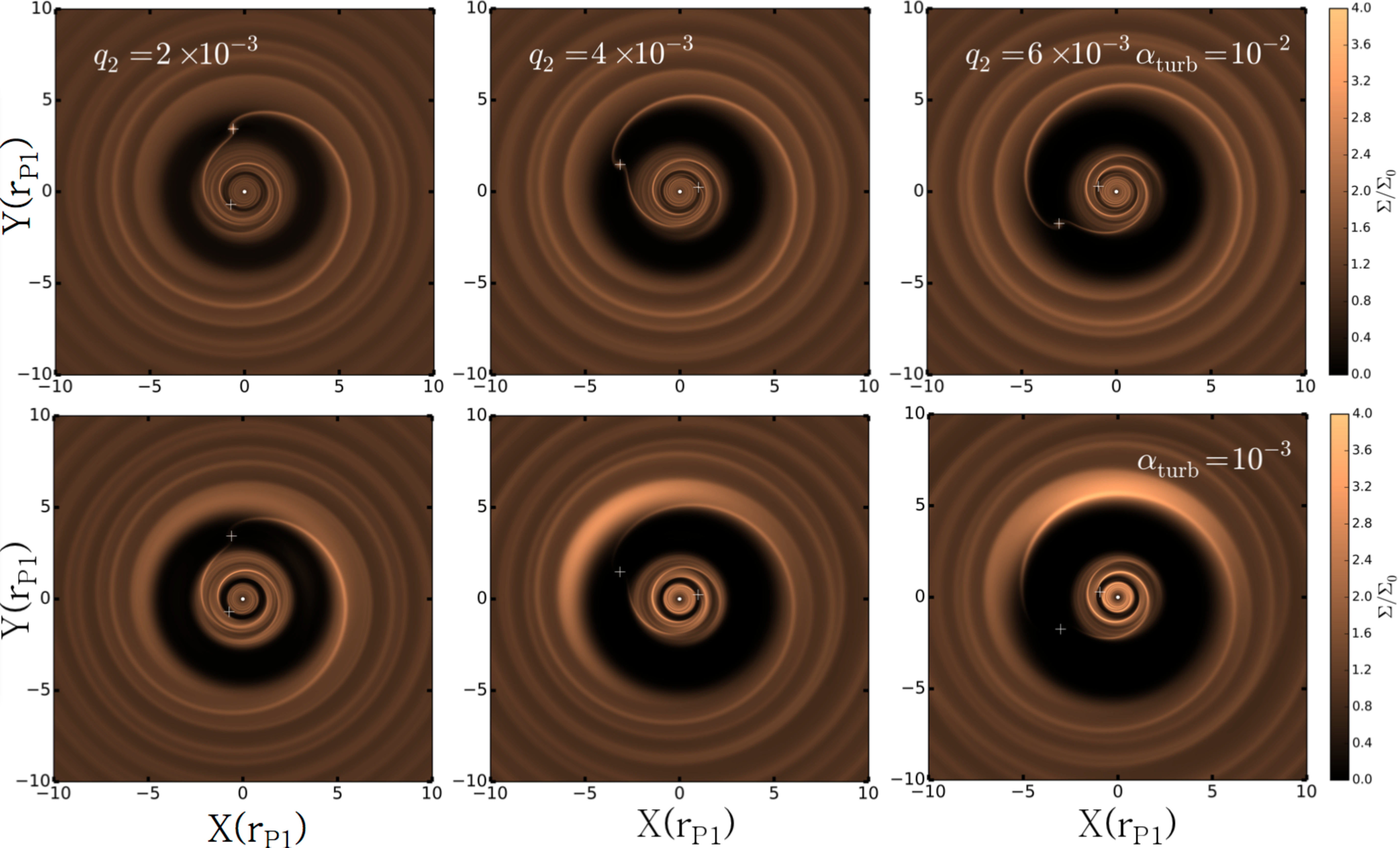}&
	\includegraphics[width=5.2cm]{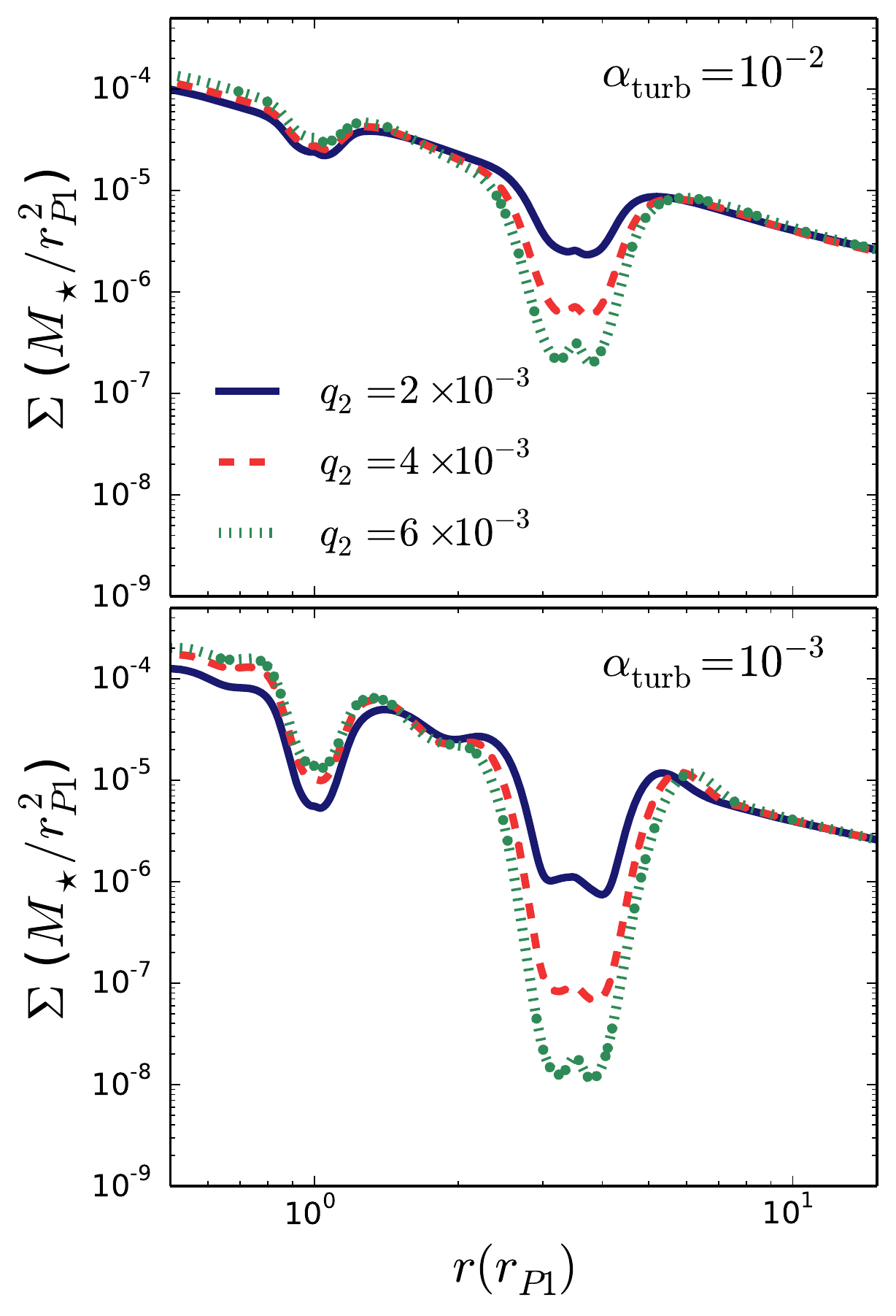}
   \end{tabular}
   \caption{Left panels: 2D gas surface density after 1000~orbits of the inner planet, when two planets with $r_{P2}/r_{P1}=3.5$  are interacting with the disk, with a planet-to-stellar mass ratio of $q_{P1}=4\times10^{-4}$ for the inner planet and   $q_{P2}=2\times10^{-3}$ (left), $q_{P2}=4\times10^{-3}$ (middle), and $q_{P2}=6\times10^{-3}$ (right), and two different values of disk viscosity $\alpha_{\rm{turb}}=10^{-2}$ (top) and $\alpha_{\rm{turb}}=10^{-3}$ (bottom). Right panels: Azimuthally averaged gas surface density from the corresponding hydrodynamical simulations and time-averaged over 200 orbits (from $1000$ to $1200$ orbits).}
   \label{hydro_results}
\end{figure*}

For vertically isothermal laminar disks, a $\sim1~M_{\rm{Jup}}$ mass planet can clear a gap \citep[e.g.][]{lin1993}. For these simulations, we consider the planet-to-central-stellar mass ratio for the inner planet ($P1$) equal to $q_{P1}=4\times10^{-4}$, which corresponds, for example,  to a $1~M_{\rm{Jup}}$ planet around a $2.5~M_{\odot}$ star.  For the outer planet ($P2$), the mass ratio is $q_{P2}=[2\times10^{-3}, 4\times10^{-3}, 6\times10^{-3}]$, which  for a stellar mass of $2.5~M_{\odot}$ implies planet masses of $5,10$ and $15~M_{\rm{Jup}}$, respectively.  These masses are taken because large separations  between the distribution of gas and millimetre dust particles  are expected for such massive planets embedded in the disk \citep{pinilla2012} and long-lived asymmetries due to eccentricity or vortex formation may exist at the edge of the gap \citep[e.g.][]{ataiee2013}. The dimensionless radial grid was taken to be logarithmically spaced from $0.1\times r_{P1}$ to $24\times r_{P1}$, and the resolution was  $n_r\times n_{\phi}=512\times587$.  We used non-reflecting conditions for the inner boundary and two values for the disk viscosity were taken for each case $\alpha_{\rm{turb}}=[10^{-3}, 10^{-2}]$. The gas surface density is initially a power-law function, $\Sigma= \Sigma_0\times (r/r_{P1})^{-1}$, and $\Sigma_0$ is taken such that the disk mass is $2.5\times10^{-3}~M_\star$.

The two planets are considered to be in fixed circular orbits, such that $r_{P2}/r_{P1}=3.5$, and they do not feel the gravity of the disk nor each other.  We adopt the position of the inner planet to be at $5$~AU, hence the outer at $17.5$~AU.  The motivation for these locations are based on the SR~21 features (although additional tests were done varying the location of these planets, finding similar conclusions). When a planet opens a gap in a disk, the outer gap edge is expected to be at most $5r_H$ for the gas and $\sim~7-10r_H$  for mm-sized particles  \citep{pinilla2012}, where the Hill radius $r_H$ is $r_H=r_p(q/3)^{1/3}$ and $q$ the planet-stellar mass ratio. Thus, under the assumptions of our simulations, a gas ring at the outer edge of the gap carved by the inner planet is expected at $\sim0.26\times r_{P1}$  from the planet (i.e. $\sim$6.3~AU from the star). In the outer regions,  when particles are trapped at the edge of the second gap, the peak of the mm-emission is expected to be at $\sim[0.61, 0.77, 0.88]\times r_{P2}$ for $q_{P2}=[2\times10^{-3}, 4\times10^{-3}, 6\times10^{-3}]$, respectively i.e. $\sim[28, 31, 33]~$AU from the star.

\begin{figure*}
 \centering
   \includegraphics[width=18.0cm]{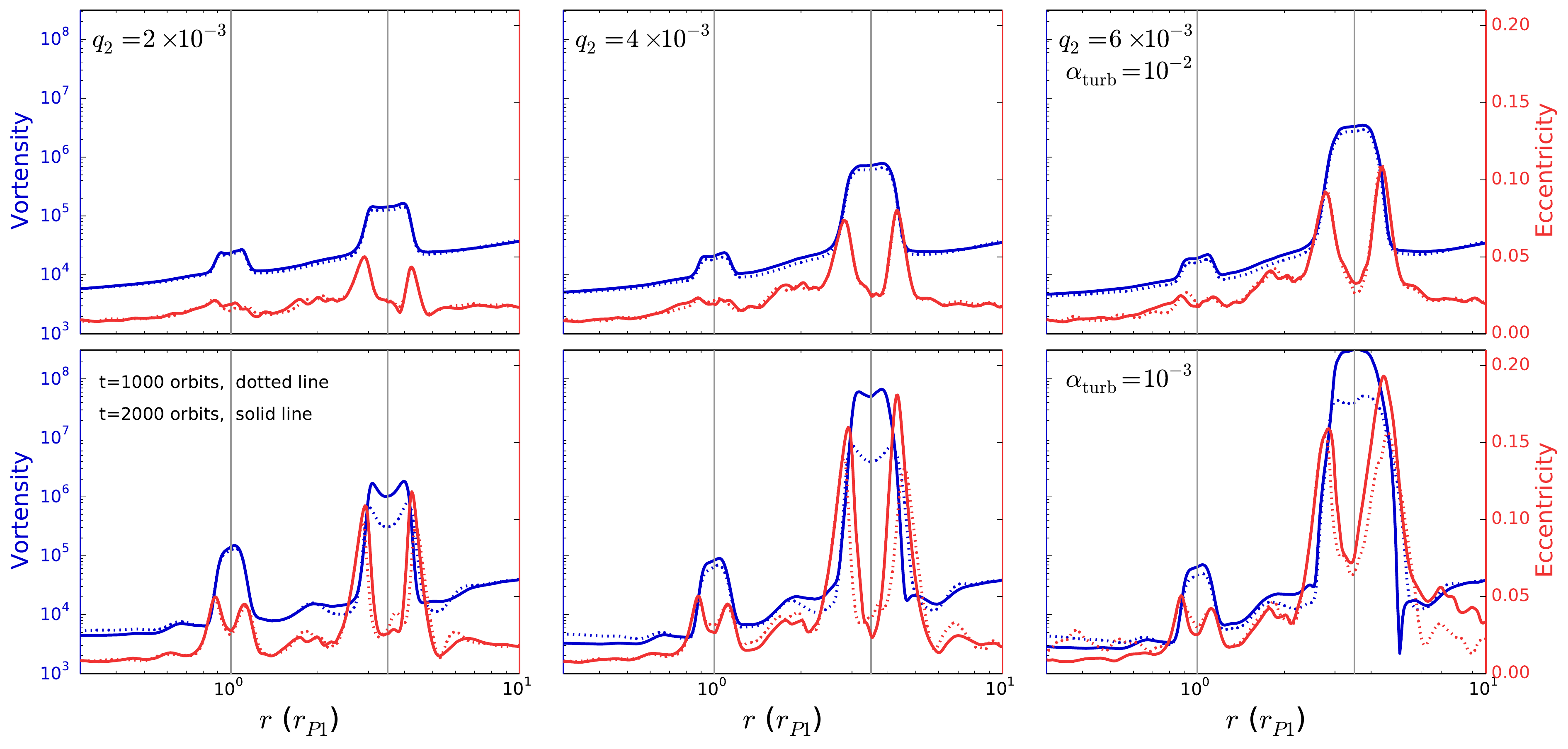}
   \caption{Azimuthally averaged vortensity $<\eta_{2\rm{D}}>_\phi$ (blue lines) and disk eccentricity (red lines) versus the disk radial extension for the hydrodynamical simulations of Fig.~\ref{hydro_results} after 1000 (dotted lines) and 2000 (solid lines) orbits. The vertical lines correspond to the positions of the two planets.}
   \label{vor_ecc}
\end{figure*}

\subsection{Dust evolution models} \label{dust_sims}

We adopted the dust evolution model explained in \cite{birnstiel2010}. This is a numerical 1D grid-based model for the global evolution of the dust surface density as introduced by \cite{brauer2008}. In this case, we consider that the dust is transported because of drag forces and mixing by turbulence.  A grid of 180 particle sizes is taken, covering grain sizes from $\sim~1\mu\rm{m}$ to $\sim~200\rm{cm}$. For the particle growth, sticking, erosion, and destructive collisional outcomes are considered. The threshold for fragmentation occurs when particles reach velocities of $v_f=[10,30]\rm{m~s}^{-1}$. Results from numerical simulations of collision between particles outside the snow line,  where grains  may have an ice mantle, indicate that the sticking efficiency is high  and that fragmentation velocities may be as high as $\sim 50~\rm{m~s}^{-1}$ \citep{wada2009}. We study the influence of planets on the dust evolution by considering the  gas surface density and velocity from the azimuthally averaged values from the hydrodynamical simulations time-averaged over 200 inner orbits (between $\sim~1000-1200$ orbits of the inner planet) as the initial conditions for the dust evolution. Dust radial drift velocities are proportional to the pressure gradient, which depends on the gas surface density, and the drag velocities depend on the gas velocities. These two components of the dust radial velocity and turbulent motion also depend on the coupling of the dust particles to the gas i.e the dust particle size and gas surface density. By taking the azimuthally averaged values, we do not consider the disk eccentricity and its effect on the final dust density distribution. In our simulations, we consider that the planets are not accreting  gas or dust. For all of the cases,  the disk mass  and the initial gas-to-dust ratio are assumed to be $M_{\rm{disk}}\sim0.006~M_{\odot}$ \citep[as][for SR~21]{andrews2011} and 100, respectively.

\subsection{Radiative transfer} \label{radiative_trans}

To generate the SEDs and images at different wavelengths, we used the 2D Monte-Carlo radiative transfer code MCMax \citep{min2009}. The vertical structure is computed considering the  particle size distribution from the dust evolution models, and assuming settling and vertical turbulent mixing consistently with $\alpha_{\rm{turb}}=[10^{-3}, 10^{-2}]$ as for the hydrodynamical and dust evolution models \citep[see][Eqs.~2 and 3]{mulders2012}. For the grain composition, we considered a mixture of  silicates ($\sim$$58\%$),  iron sulphide ($\sim$$18\%$), and carbonaceous  ($\sim$$24\%$) grains as in \cite{dejuanovelar2013}. For the inclination and distance to the source, we assume $i =15^\circ$ and $d=135$~pc. All parameters are summarised in Table~\ref{table_inputs}.

\section{Results of numerical simulations}     \label{results}
In this section, we present the main results from the hydrodynamical simulations, together with the dust evolution models.

\subsection{Planet-disk interactions} 

The left panels of  Fig.~\ref{hydro_results} show  2D gas surface density after 1000~orbits of the inner planet (hereafter used as the time unit for the hydrodynamical results) when two planets  interact with the disk. When the  planet-to-stellar mass ratio of the outer planet increases, the corresponding gap is wider and deeper because of the higher angular momentum transfer, as shown  by several authors \citep[see][for an extensive review]{kley2012}. The effect of lowering the viscous torque, by assuming a lower value for $\alpha_{\rm{turb}}$, also has an effect on the gap shape \citep[e.g.][]{crida2006}. This is particularly true in the outer edge of the second gap, where a vortex is formed as in the case of a massive single planet interacting with a disk of moderate viscosity \citep[e.g.][]{ataiee2013, fu2014}. The right panels of Fig.~\ref{hydro_results} illustrate the azimuthally averaged gas surface density from the corresponding hydrodynamical simulations, time averaged over 200 orbits (between $1000-1200$ orbits). This explicitly shows the effect of the planet mass and viscosity on the gap shape. These gas surface density profiles are the initial conditions for the dust models and we assume that these time-averaged profiles remain constant during the dust evolution.

\begin{figure*}
 \centering
   \includegraphics[width=18.0cm]{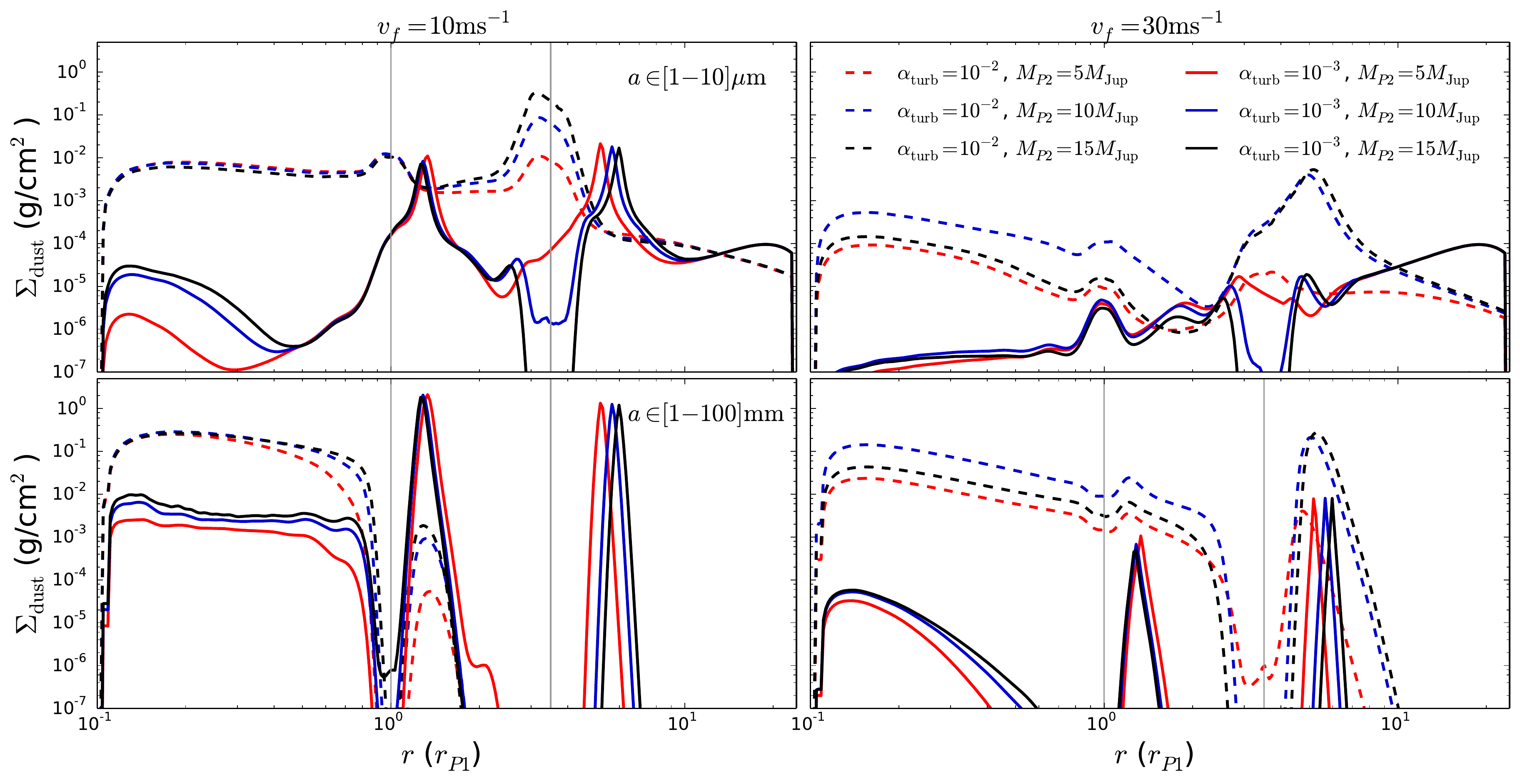}
   \caption{Dust surface density for micron- (upper)  and millimetre-sized particles (bottom), assuming two planets interacting with the disk at 5~AU and 17.5~AU and after 1~Myr of dust evolution. Two different thresholds are considered for the fragmentation velocity of particles $v_f=10\rm{m~s}^{-1}$  (left) and $v_f=30\rm{m~s}^{-1}$ (right). The vertical lines correspond to the positions of the two planets.}
   \label{dust_distribution}
\end{figure*}

The interaction of a planet(s) with a disk induces gas asymmetries, which may be the result of disk eccentricity \citep[e.g.][]{papaloizou2001, dangelo2006, kley2006, Hosseinbor2007, lubow2010}, vortex formation due to Rossby wave instability (RWI) \citep[e.g.][]{li2000, li2005, lin2012, lin2014, meheut2012}, or a mix of both. The Rossby wave instability develops at the local minima of potential vorticity or vortensity ($\eta$), triggering anticyclonic vortices. The vorticity is the tendency of a fluid to locally rotate around a central point, thus for a fluid with $u$ velocity, vorticity is defined as $\mathbf{\nabla} \times \mathbf{u}$, and the vortensity is the vorticity weighted by the density. If the minimum of $\eta$ is deep, the RWI is strong and therefore the resulting vortex is also strong. In 2D, vortensity $\eta_{2\rm{D}}$ is defined as the ratio of the vertical component of the vorticity i.e. $\hat{z} \cdot \mathbf{\nabla} \times \mathbf{u}$  to surface density \citep[e.g.][]{lovelace1999, lin2014}.

Figure~\ref{vor_ecc} illustrates the azimuthally averaged vortensity and disk eccentricity \citep[calculated as in][]{kley2006} versus the disk radial extension for the hydrodynamical simulations of Fig.~\ref{hydro_results}, after 1000 and 2000 orbits. Both vortensity and eccentricity increase with higher planet-to-stellar mass ratio as well as for lower disk viscosity ($\alpha_{\rm{turb}}$). Nonetheless, the vortensity is almost flat at the edges of these gaps for any case of $\alpha_{\rm{turb}}=10^{-2}$, implying no long-lived anticyclonic vortices. For $\alpha_{\rm{turb}}=10^{-3}$, there are local vortensity minima at the outer edge of the gap carved by the second planet when $q_{P2}=4\times10^{-3}$ and $q_{P2}=6\times10^{-3}$, implying potential RWI that may trigger the formation of anticyclonic vortices, which can survive over thousands of orbits (Fig.~\ref{hydro_results}). For these two cases, the azimuthal asymmetries are a combination of eccentricity and vortex formation.  After 2000 orbits, the maximum values of the eccentricity for $q_{P2}=[2\times10^{-3}, 4\times10^{-3}, 6\times10^{-3}]$  are  $e_{\rm{max}}\sim~[0.05, 0.08, 0.11]$ for  $\alpha_{\rm{turb}}=10^{-2}$ and $e_{\rm{max}}\sim~[0.12, 0.18, 0.20]$  for  $\alpha_{\rm{turb}}=10^{-3}$. 

\subsection{Dust density distribution} 

In Fig.~\ref{dust_distribution}, we show the dust surface density for micron- and millimetre-sized particles after 1~Myr of dust evolution, assuming  the gas surface density from the hydrodynamical simulations as the initial condition. Two different thresholds are considered for the fragmentation velocity of particles $v_f=10~\rm{m~s}^{-1}$ and $v_f=30~\rm{m~s}^{-1}$. The distribution of small and large grains is strongly affected by the shape of the gap i.e. the initial gas surface density, the turbulent parameter  $\alpha_{\rm{turb}}$, and the fragmentation velocity $v_f$.

\begin{table*}
\caption{Results for $M_{P1}=1~M_{\rm{Jup}}$ and $v_f=10~\rm{m~s}^{-1}$, after 1~Myr of evolution}
\centering   
\tabcolsep=0.05cm                   
\subtable[$M_{P2}=5~M_{\rm{Jup}}$]{
\begin{tabular}{c|cc}       
\textbf{{\small Potential Feature}}& $\alpha_{\rm{turb}}=10^{-2}$&$\alpha_{\rm{turb}}=10^{-3}$\\
\hline
{\scriptsize Radial trapping in}&&\\
{\scriptsize pressure maxima}&$\times$&$\checkmark$\\
\hline
{\scriptsize No radial gaps of}&&\\
{\scriptsize micron-sized particles}&$\checkmark$&$\checkmark$\\
\hline
{\scriptsize Effective}&&\\
{\scriptsize fragmentation}&$\checkmark$&$\checkmark$\\
\hline
\end{tabular}    
}
\subtable[$M_{P2}=10~M_{\rm{Jup}}$] {
\begin{tabular}{cc}       
$\alpha_{\rm{turb}}=10^{-2}$&$\alpha_{\rm{turb}}=10^{-3}$\\
\hline
\\
$\times$&$\checkmark$\\
\hline
\\
$\checkmark$&$\times$\\
\hline
\\
$\checkmark$&$\checkmark$\\
\hline
\end{tabular}    
}
\subtable[$M_{P2}=15~M_{\rm{Jup}}$] {
\begin{tabular}{cc}       
$\alpha_{\rm{turb}}=10^{-2}$&$\alpha_{\rm{turb}}=10^{-3}$\\
\hline
\\
$\times$&$\checkmark$\\
\hline
\\
$\checkmark$&$\times$\\
\hline
\\
$\checkmark$&$\checkmark$\\
\hline
\end{tabular}    
}
\label{results1}
\end{table*}

\begin{table*}
\caption{Results  for $M_{P1}=1~M_{\rm{Jup}}$ and  $v_f=30~\rm{m~s}^{-1}$, after 1~Myr of evolution}
\centering   
\tabcolsep=0.05cm                   
\subtable[$M_{P2}=5~M_{\rm{Jup}}$]{
\begin{tabular}{c|cc}       
\textbf{{\small Potential Feature}}& $\alpha_{\rm{turb}}=10^{-2}$&$\alpha_{\rm{turb}}=10^{-3}$\\
\hline
{\scriptsize Radial trapping in}&&\\
{\scriptsize pressure maxima}&$\times$&$\checkmark$\\
\hline
{\scriptsize No radial gaps of}&&\\
{\scriptsize micron-sized particles}&$\checkmark$&$\checkmark$\\
\hline
{\scriptsize Effective}&&\\
{\scriptsize fragmentation}&$\checkmark$&$\times$\\
\hline
\end{tabular}    
}
\subtable[$M_{P2}=10~M_{\rm{Jup}}$] {
\begin{tabular}{cc}       
$\alpha_{\rm{turb}}=10^{-2}$&$\alpha_{\rm{turb}}=10^{-3}$\\
\hline
\\
$\checkmark$&$\checkmark$\\
\hline
\\
$\checkmark$&$\checkmark$\\
\hline
\\
$\checkmark$&$\times$\\
\hline
\end{tabular}    
}
\subtable[$M_{P2}=15~M_{\rm{Jup}}$] {
\begin{tabular}{cc}       
$\alpha_{\rm{turb}}=10^{-2}$&$\alpha_{\rm{turb}}=10^{-3}$\\
\hline
\\
$\checkmark$&$\checkmark$\\
\hline
\\
$\checkmark$&$\checkmark$\\
\hline
\\
$\checkmark$&$\times$\\
\hline
\end{tabular}    
}
\label{results2}
\end{table*}

Because of the enhancement of the gas surface density at the outer edges of the gaps, there are pressure maxima and therefore preferential regions for particles to drift. Nevertheless, turbulence can push particles out from these pressure bumps. The balance between the turbulent mixing and the strength of the positive pressure gradient at these locations determines if grains can be trapped or not. Indeed, in the regions where the radial drift is reduced, the relative velocities between particles are dominated by turbulent motion, and the maximum particle size before they fragment ($a_{\mathrm{max}}$) is parametrised as in \cite{birnstiel2009}

\begin{equation}
a_{\mathrm{max}}\propto \frac{\Sigma_g}{\alpha_{\mathrm{turb}}\rho_s}\frac{v_f^2}{c_s^2},
\label{a_max}
\end{equation}

\noindent where $\Sigma_g$ is the gas surface density and $\rho_s$ the volume density of a grain (of the order of 1g~cm$^{-3}$). Thus, if the maximum particle size in pressure bumps results in bodies that are more affected by the turbulent mixing than radial drift, particle trapping would not happen. 

\begin{figure}
 \centering
   \includegraphics[width=9.0cm]{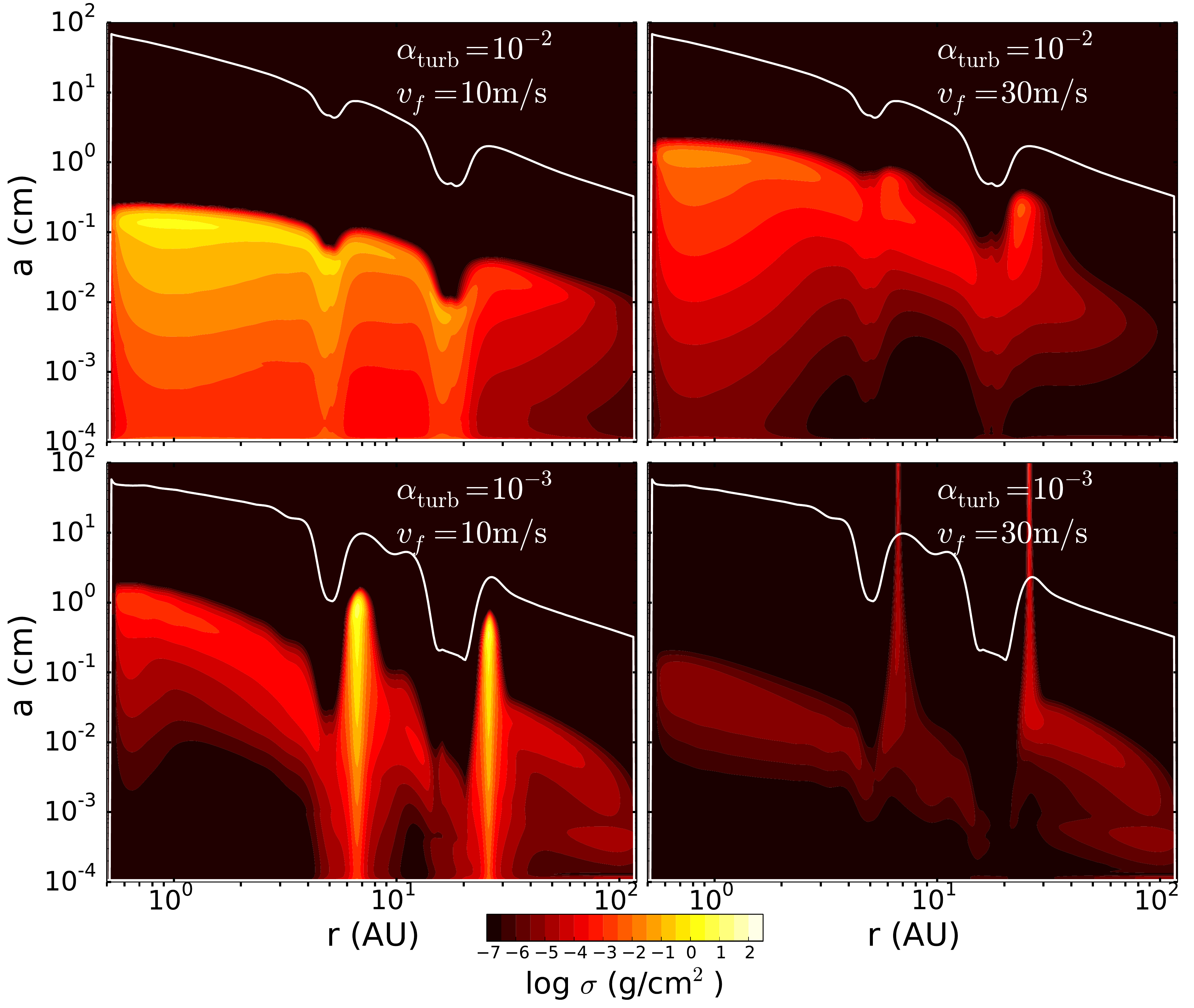}
   \caption{Dust density distribution  at 1~Myr of evolution when two planets are interacting with a disk at 5~AU  ($1M_{\rm{Jup}}$) and at 17.5~AU ($5M_{\rm{Jup}}$) with  $\alpha_{\rm{turb}}=10^{-2} (top), v_f=10\rm{m~s}^{-1}$ (left) and $\alpha_{\rm{turb}}=10^{-3} (bottom), v_f=30\rm{m~s}^{-1}$ (right). The solid line corresponds to  the particle size for which particles experience the highest radial drift, which is directly proportional to $\Sigma_g$ (Eq.~\ref{a_max}).}
   \label{dust_distribution2}
\end{figure}

The simulations with $\alpha_{\rm{turb}}=10^{-2}$ and $v_f=10\rm{m~s}^{-1}$, show how turbulence prevails over the potential trapping of particles and there is no dust accumulation at the outer edge of the gap carved by the  second planet, independent of $M_{P2}$ (Fig.~\ref{dust_distribution}).  In this case, the increment of the dust surface density for small particles at the location of the planets happens because, in the outer regions, large grains fragment and the  small dust co-moves with gas radial velocities without being trapped. The inward motion of dust particles from the outer edge through the gap creates a pile-up of small particles close to the planet location because of fragmentation, where the maximum size of particles ($a_{\rm{max}}$)  is lower than outside the gap (because of the lower gas surface density, see Eq.~\ref{a_max}). When the fragmentation velocity is higher, particles can reach larger sizes (Eq.~\ref{a_max}), leading to  trapping of dust particles at pressure maxima, except for the case of $M_{P2}=5~M_{\rm{Jup}}$, where the positive pressure gradient is not high enough to stop the rapid inward drift of particles. For these high fragmentation velocities ($v_f=30\rm{m~s}^{-1}$) and viscosity ($\alpha_{\rm{turb}}=10^{-2}$), micron-sized particles are distributed all over the disk, with an enhancement at the location of the outer trap due to the continous fragmentation of large grains. 

In contrast, in the case of $\alpha_{\rm{turb}}=10^{-3}$, there is effective trapping of millimetre-grain particles in both gap outer edges, and independent of the fragmentation velocities. When $v_f=10\rm{m~s}^{-1}$, the constant fragmentation, by turbulent motion of large grains enriches the disk of smaller grains. In particular, this fragmentation occurs at the location of the pressure traps, which also leads to an increment of the surface density of micron-sized particles \citep[see][]{andrews2014}. When $\alpha_{\rm{turb}}=10^{-3}$ and $v_f=30\rm{m~s}^{-1}$, fragmentation is unlikely to happen, and there are fewer micron-sized particles in the disk. For these cases, the dust surface density of mm-grains  (Fig.~\ref{dust_distribution}) is dominated by the very large grains i.e. $a>10$~mm.

The location and shape of the corresponding millimetre rings depend on the planet mass and viscosity. For a more massive outer planet, the mm dust particles  accumulate further away \citep{pinilla2012}. The width is affected by the values taken for  $\alpha_{\rm{turb}}$, leading to a wider ring when turbulent mixing is higher.  The amount of small and large grains in the inner disk ($r<5$~AU), also depends on the turbulence strength and potential trapping. For instance, by comparing the cases of $v_f=10\rm{m~s}^{-1}$ and the different values of $\alpha_{\rm{turb}}$, when trapping occurs ($\alpha_{\rm{turb}}=10^{-3}$), there is a much smaller amount of micron-sized particles outside the pressure traps (the same happens  for the same $\alpha_{\rm{turb}}$, but with higher fragmentation velocity).

Figure~\ref{dust_distribution2} illustrates in more detail the dust distribution of all grain sizes and how dust grains are trapped or not for each condition, for the particular case of $q_{P2}=2\times10^{-3}$ (i.e. 5$M_{\rm{Jup}}$ around a $2.5~M_{\odot}$ star). When fragmentation does not occur (as in the case of $v_f=30\rm{m~s}^{-1}$ and $\alpha_{\rm{turb}}=10^{-3}$), particles continue growing to the largest size considered in the simulations (200~cm), leading to an optically thin disk at all wavelengths. 

\begin{figure*}
 \centering
   \includegraphics[width=18.0cm]{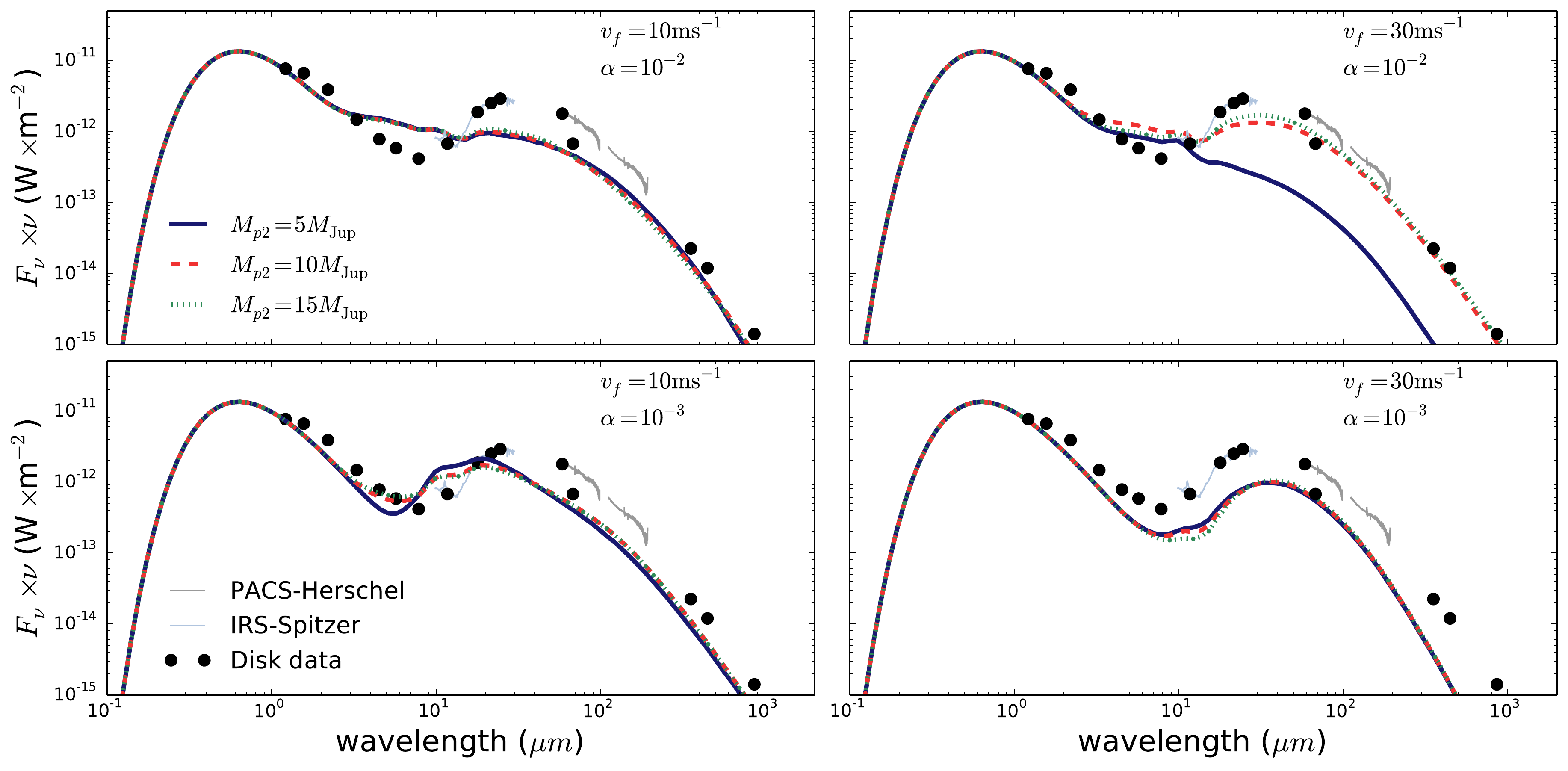}
   \caption{Spectral energy distribution (SED) resulting from the radiative transfer modelling, assuming  grain size distribution after 1~Myr of evolution when two planets interact with the disk, and assuming different values for the disk viscosity $\alpha_{\rm{turb}}$ and fragmentation velocity of particles $v_f$. The disk data are for SR~21 and are taken from \cite{brown2007, follette2013} and PACS (DIGIT PI: N.~Evans).}
   \label{SEDs}
\end{figure*}

From the performed simulations, three cases allow radial trapping of particles at the outer edges of the carved gaps, a smooth radial distribution of micron-sized particles i.e. no total filtration of particles, and an effective fragmentation of dust grains i.e. no continuous growth to planetesimal-sized objects. These cases are: $\alpha_{\rm{turb}}=10^{-3}$, $v_f=10\rm{m~s}^{-1}$ and $M_{P2}=5~M_{\rm{Jup}}$, and $\alpha_{\rm{turb}}=10^{-2}$, $v_f=30\rm{m~s}^{-1}$ and $M_{P2}=\{10,15\}~M_{\rm{Jup}}$. Tables~\ref{results1} and \ref{results2} summarise the main findings of the simulations done for this paper.  To ensure the trends described in this paper, we repeated some of the simulations considering higher disk mass (by a factor of 2), larger distance between planets ($r_{P2}/r_{P1}=4.5$), and smaller initial grain size ($0.1~\mu$m), and found similar results. 

In our models, we assume that the planets do not migrate and that they do not accrete gas or dust. Migration can create resonances, affect the eccentricity, and change the location of the pressure traps, which can affect the final distributions of dust. For instance, the inner planet partially opens a gap (Fig.~\ref{hydro_results}), and therefore it can  experience a rapid migration, inhibiting the particle trapping at the outer edge of its gap. The masses considered for the outer planet are high enough for a slow migration, allowing particles to be trapped at the outer edges of the corresponding gaps. These particle concentrations would move while the planet is migrating. When planets do not migrate, resonances between the planets are not expected because their separation is large.  In this case, the maximum value found for the eccentricity between the two planets is $\sim0.15$ (Fig.~\ref{vor_ecc}). This eccentricity is not high enough to overlap the gaps, implying that for all cases there is a long-lived gas ring between planets that, depending on different parameters (Tables ~\ref{results1} and \ref{results2}), may or may not trap particles. Moreover, accretion of gas and dust onto the planet may also change the gas and dust distributions close to the planet orbits \citep[e.g.][]{owen2014}.

\section{Comparison with observations }     \label{observations}

In this section, we present the SEDs and synthetic sub-millimetre and polarised infrared images computed from the models. 

\subsection{Spectral energy distributions}

Figure~\ref{SEDs} shows the SEDs from radiative transfer simulations for all the cases introduced in Sect.~\ref{results}, when two planets interact with the disk, different values for the disk viscosity $\alpha_{\rm{turb}}$, and fragmentation velocity of particles $v_f$. We consider the grain size distribution from the dust evolution models after 1~Myr of evolution. For comparison,  in Fig~\ref{SEDs}, we include data from \cite{follette2013}, the PACS (DIGIT data) and IRS spectra \citep{brown2007} for the case of  SR~21 disk.

\begin{figure*}
 \centering
  \begin{tabular}{cc}  
   \includegraphics[width=9.0cm]{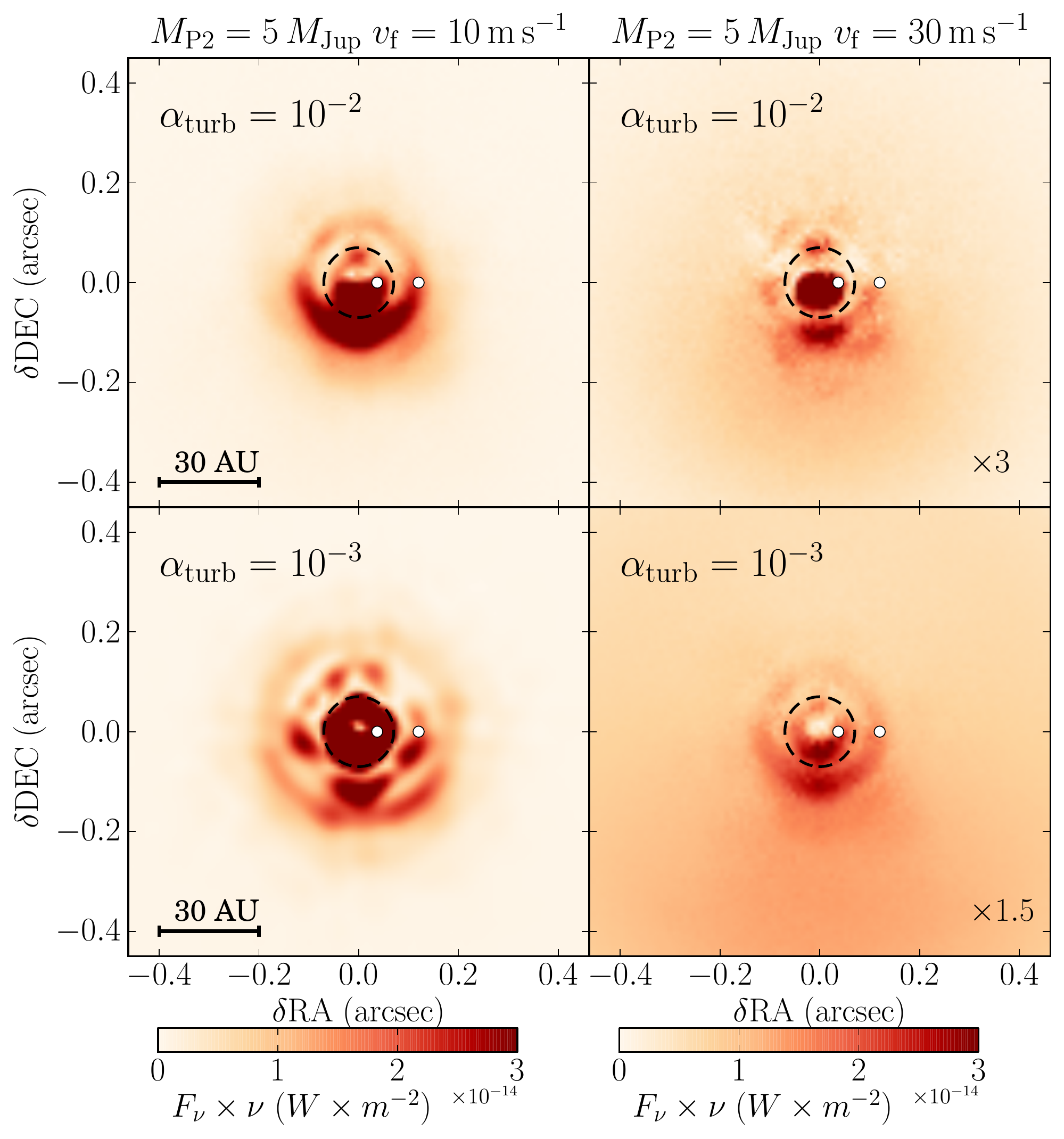}&
   \includegraphics[width=9.0cm]{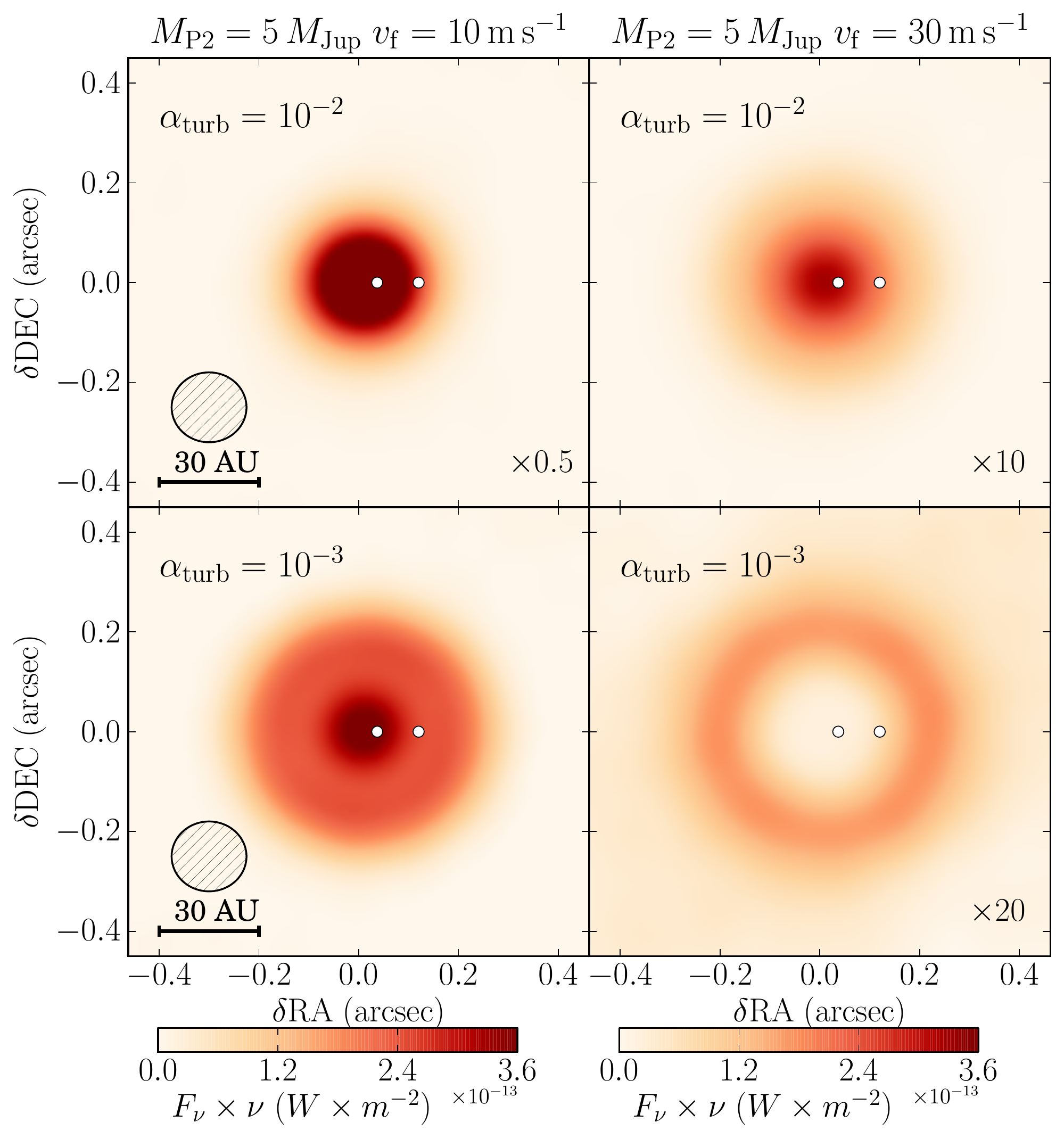}
     \end{tabular}
   \caption{Synthetic H-band HiCIAO (left panels) and ALMA Band 7  ($850~\mu$m)  (right panels) observations for the models of an outer planet $M_{P2}=5M_{\rm{Jup}}$, and each value of the fragmentation velocity $v_f$ and turbulence parameter  $\alpha_{\rm{turb}}$. The white dashed line in the left panels indicates the inner $0\arcsec.07$ of HiCIAO images masked in the \citet{follette2013} observations of SR~21. In the right panels, the size of the ALMA Cycle 2 beam is indicated by a stripped circle.}
   \label{images}
\end{figure*}

With a fragmentation velocity of  $v_f=10\rm{m~s}^{-1}$ and  $\alpha_{\rm{turb}}=10^{-2}$, there is no effective trapping of particles in the outer pressure maxima for any $M_{P2}$ (Fig.~\ref{dust_distribution}) and  no dust depleted cavity, thus there is no characteristic deficit  of emission at near/mid-infrared  as seen in transition disks. In addition, the emission at far-infrared and mm-wavelength is under-predicted.  When  $\alpha_{\rm{turb}}=10^{-3}$ and trapping of particles is possible, the mid-infrared emission decreases compared to the cases of $\alpha_{\rm{turb}}=10^{-2}$ because of the cleared cavity. In the cases of $\alpha_{\rm{turb}}=10^{-3}$, the planet mass slightly influences the SEDs. For instance, when $M_{P2}=5M_{\rm{Jup}}$, where there is less dust filtration at the corresponding gap (and therefore a larger amount of micron-sized particles at these locations, but less in the inner disks),  there is lower near-infrared emission, but larger mid-infrared flux. Between $M_{P2}=10M_{\rm{Jup}}$ and $M_{P2}=15M_{\rm{Jup}}$, there is almost no difference in the SEDs profiles because dust filtering is similar. 

When $v_f=30~\rm{m~s}^{-1}$, $\alpha_{\rm{turb}}=10^{-2}$, and $M_{P2}=5M_{\rm{Jup}}$, there is no trapping and not enough dust to produce mid- and far-infrared emission (Fig.~\ref{dust_distribution}). Models with $M_{P2}=10M_{\rm{Jup}}$ and $M_{P2}=15M_{\rm{Jup}}$ allow many large grains in the outer part of the disk and therefore there is a higher far-infrared and millimetre emission. However, there are also a large amount of dust particles within the cavity, so no strong deficiency of the mid-infrared emission is generated. For these cases, the fact of having higher turbulence helps to increase the scale-height of the disk and the amount of dust in the upper layers, implying a larger amount of reprocessed stellar light. In the case of no fragmentation  ($v_f=30\rm{m~s}^{-1}$ and  $\alpha_{\rm{turb}}=10^{-3}$), particles grow to very large sizes, causing  the disk to be optically thin at most wavelengths and producing a debris-disk type SED.  

The best models found in Sect.~\ref{results}, which allow trapping of millimetre particles, effective fragmentation, and continuous distribution of micron-sized particles, are suitable for reproducing the characteristic SEDs of transition disks. The models of $\alpha_{\rm{turb}}=10^{-3}$, $v_f=10\rm{m~s}^{-1}$, and $M_{P2}=5~M_{\rm{Jup}}$ may under-predict the typical far-infrared and mm-emission. Assuming a more massive disk or/and higher dust-to-gas ratio could increase the emission at mm-wavelengths. Some of the simulations were repeated with a higher disk mass (by a factor of 2), without significantly changing the resulting trends of the SEDs.  In this case, the lack of far-infrared flux may be a consequence of a small disk scale height at the outer disk rim, where $h(r)$ could be larger than in the inner disk \citep[e.g.][]{maaskant2013}. On the other hand, the models of $\alpha_{\rm{turb}}=10^{-2}$, $v_f=30\rm{m~s}^{-1}$, and $M_{P2}=\{10,15\}~M_{\rm{Jup}}$ reproduce a better fitting of the far-infrared and mm-emission, however, the dip at $10~\mu$m is less evident because of the homogeneous distribution of  grains in the inner part of the disk (Fig~\ref{dust_distribution}).

\subsection{Polarised near-infrared emission and millimetre maps}

In Fig.~\ref{images} we show synthetic HiCIAO H-band polarised intensity and ALMA Band 7  ($850~\mu$m) intensity  images, for the models of an outer planet $M_{P2}=5~M_{\rm{Jup}}$ and each value of the fragmentation velocity $v_f$ and turbulence parameter  $\alpha_{\rm{turb}}$, assuming an inclination angle of $i=15^\circ$ and a distance of 135~pc. The HiCIAO images are obtained convolving full resolution $1.6\mu m$ images with a measured HiCIAO point spread function (PSF) in H-band taken from the ACORNS-ADI SEEDS Data Reduction Pipeline software \citep{brandt2013}. The ALMA images are obtained with the CASA simulator (version 4.1.0) assuming Cycle~2 capabilities and one hour of total observation time (antenna configuration that gives a beam of $\sim~0.15\arcsec\times0.14\arcsec$ \footnote{Antenna configuration C34-7 from \url{http://almascience.eso.org/documents-and-tools}}). These images trace opposite ends of the dust particle size distribution i.e.\,polarised intensity images at short wavelengths trace small (1-10~$\mu$m) dust grains, while sub-millimetre observations trace large (1-10~mm) grains.

Polarised intensity observations of the cases of $M_{P2}=5~M_{\rm{Jup}}$, $v_f=\{10, 30\}\rm{m~s}^{-1}$ and $\alpha_{\rm{turb}}=10^{-2}$, in which there is no trapping of dust particles and an enhancement of the dust surface density of small particles close to the position of the planets, show ring structures at those locations. Note that the full resolution images are azimuthally symmetric and therefore the speckle features seen in the H-band images are due to the shape of the HiCIAO PSF and do not correspond to morphological features of the disk. The images show an asymmetry in brightness between the far (top) and near (low) regions of the disk which is a result of the forward scattering of starlight by the small dust particles and the fact that we are looking at an inclined disk (i.e. a face-on disk would not show such asymmetry in brightness). The ALMA images for these two cases do not show any gap because the disk is depleted of large particles at the outer parts of the disk.

When there is trapping, and there is still a population of small-dust particles distributed over the disk (case of $M_{P2}=5~M_{\rm{Jup}}$ $v_f=10\rm{m~s}^{-1}$ and $\alpha_{\rm{turb}}=10^{-3}$), the intensity should be distributed over the whole disk with a smooth depletion between the two planets (Fig~\ref{dust_distribution}). The PSF features mask this region, however, it is still possible to detect a ring of small particles that accumulate at the location of the outer particle trap. The ALMA image in this case shows a slight cavity between the two planets, but the two traps are too close for the gap to be resolved. 

For $\alpha_{\rm{turb}}=10^{-3}$ and $v_f=30\rm{m~s}^{-1}$, the HiCIAO image reflects a lack of small particles, when fragmentation and trapping are unlikely. The ALMA image shows a very faint ring of millimetre particles created at the outer edge of the gap, where grains are continuously growing to the allowed maximum size (Fig~\ref{dust_distribution2}). Note that the flux in both HiCIAO and ALMA images for all simulations with $v_f=30\rm{m~s}^{-1}$ has been multiplied by a factor of $1.5$ and $20$, respectively, in order to show the features, but these features are therefore difficult to detect.

In the cases of Fig~\ref{images}, to reproduce a disentanglement between large and small grains, the most feasible case is when $v_f=10\rm{m~s}^{-1}$, $\alpha_{\rm{turb}}=10^{-3}$, and $M_{P2}=5M_{\rm{Jup}}$, where micron-sized particles are located over most of the disk ($r>5$AU), and mm-dust particles are trapped in narrow rings at $\sim~$7AU and $\sim~$27AU. In this case,  ALMA in Cycle~2 is not capable of resolving the gap between the two particle traps, however, a future configuration of 50 antennas can resolve this gap. 

\section{Discussion}     \label{discussion}
In this section, we focus the discussion on comparing the previous results to the case of  the disk of  SR~21. For this target, we consider that the observed gas-ring detected with CRIRES at 7~AU is the result of the inner planet opening a gap, which is located at 5~AU. The mm-cavity at $\sim$36~AU arises from the interaction with the additional outer massive planet at  $\sim~15-20$~AU.

\subsection{The asymmetry}

To fit the SR~21  ALMA observations,  \cite{perez2014} fit a Gaussian  profile in the radial and azimuthal direction following the vortex morphology introduced by \cite{lyra2013} and compare it to a Gaussian radial ring \citep[see also][]{birnstiel2013}. The best-fit model corresponds to a vortex, located at $\sim46$~AU from the central star ($\sim$~10~AU further than the fitting of a ring), centre at $\sim178^\circ$ (measured from east to north), with a radial and an azimuthal width of $14.4$~AU and $40.4$~AU, respectively. The best fit corresponds to a vortex with aspect ratio ($\chi$)  of 2.8. Assuming particles of $a=1$mm size, with $\rho_s=1~\rm{g}~\rm{cm}^{-3}$, and $\Sigma_{g}=10~\rm{g}~\rm{cm}^{-2}$, their turbulent velocities within the vortex are about 16\% of the sound speed. These turbulent velocities $v_{\rm turb}$ \citep[$v_{\rm turb}\propto\sqrt{\alpha_{\rm{turb}}}c_s$, see][]{ormel2007} correspond to a diffusion parameter $\alpha_{\rm{turb}}$ of around $\sim~2\times10^{-2}$ with temperatures of $\sim$30-35~K at the vortex location. Although magneto-rotational instability simulations suggest values of  $\alpha_{\rm{turb}}$ to be in the range of $10^{-3}-10^{-2}$ \citep[e.g.][]{johansen2005}, steady-state vortices are unlikely to form and be long-lived at the outer edge of the gap if these values for $\alpha_{\rm{turb}}$ are assumed for the whole disk (Fig.~\ref{vor_ecc}). However, this high turbulence might be only at the vortex location and created by the vortex itself. On the other hand, their  best fit for the vortex width ($\sigma_{r, V}\sim~14.4$~AU) seems to be  too wide as compared to the scale height of the disk  ($\sim~2$~AU, considering a temperature of  $T\sim35$~K at the peak of the emission $\sim$~40~AU). To sustain  long-lived vortices in protoplanetary disks, the gas velocities must remain sub-sonic and the radial width of a vortex therefore cannot be much larger than the scale height of the disk \citep[see e.g.][]{barranco2005}.  Because of the ALMA beam size of $\sim$27~AU, the vortex width is still not resolved and we cannot disprove the vortex-scenario.

\begin{figure}
\centerline{
   \includegraphics[width=10.0cm]{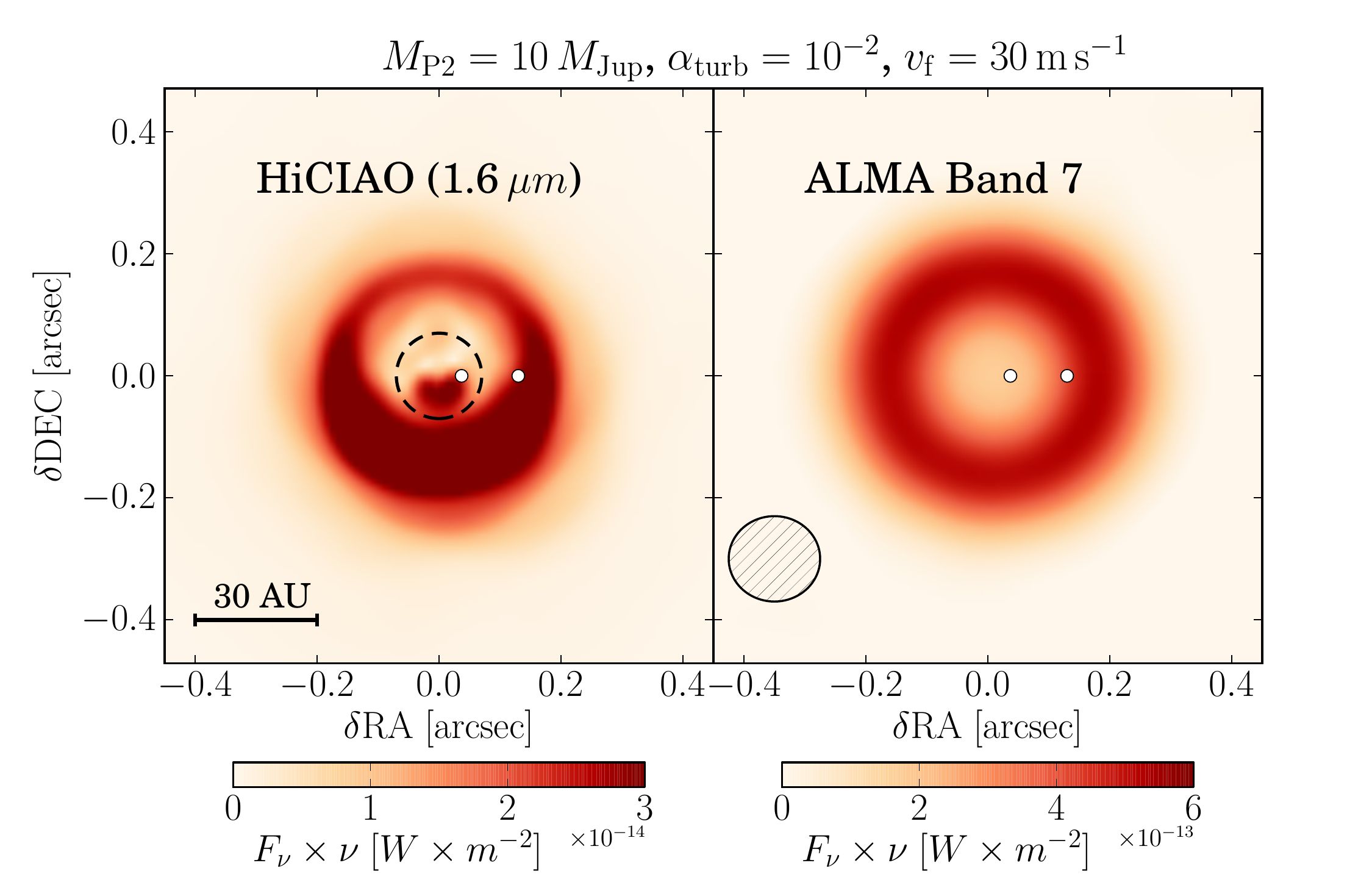}}
   \caption{Synthetic H-band HiCIAO (left panel) and ALMA Band 7 ($850~\mu$m) (right panel) observations for the case of an outer planet of $M_{P2}=10M_{\rm{Jup}}$, $v_f=30\rm{m~s}^{-1}$ and $\alpha_{\rm{turb}}=10^{-2}$. White dashed circle indicates the inner $0\arcsec.07$ of HiCIAO images masked in the \citet{follette2013} observations of SR~21. Stripped circle in right panel indicates the size of the ALMA Cycle 2 beam.}
   \label{images_10Mjup}
\end{figure}

An alternative scenario for this asymmetry is disk eccentricity.  Eccentric gaps have been observed for some transition disks. For example, using NIR high contrast imaging, \cite{thalmann2014} report an eccentric gap ($e\approx0.3$) for the LkCa~15 disk. Contrary to an anticyclonic vortex, over-densities in an eccentric disk do not  trap particles.  Because of the low gas velocities at the aphelion compared to the perihelion, gas, and dust densities are expected to be higher at the aphelion, creating asymmetries that could be also detected with high resolution and sensitivity observations of CO line profiles and continuum \citep{regaly2011, hsieh2012, ataiee2013, regaly2014}. The contrast of the gas and dust densities are expected to be similar, which is very different from the case of a vortex where the contrast of the millimetre-sized dust density can be much higher than in the gas \citep{birnstiel2013}. For example, with a gas surface contrast of $\sim~4$ due to the vortex (case of $\alpha_{\rm{turb}}=10^{-3}$ and $q_{P2}=6\times10^{-3}$) dust contrasts of more than $100$ are expected. 

From the hydrodynamical simulations in Sect.~\ref{results}, both eccentricity and vortex formation are expected in different cases (Fig.~\ref{vor_ecc}). Eccentricity dominates any possible asymmetry when $\alpha_{\rm{turb}}=10^{-2}$, with a maximum value of $e\sim0.1$ for $q_{P2}=6\times10^{-3}$ or $M_{P2}=15M_{\rm{Jup}}$ and creates a slight gas surface density contrast in the azimuthal direction. However,  spiral arms created by the outer planet may strengthen those asymmetries (Fig.~\ref{hydro_results}).  For $\alpha_{\rm{turb}}=10^{-3}$, potential asymmetries result from the combination of vortex formation and disk eccentricity, besides the case of $q_{P2}=2\times10^{-3}$ or $M_{P2}=5M_{\rm{Jup}}$ for which the formed vortex dissipates at longer time-scales.

\subsection{Spatial segregation of small and large grains}

According to the dust distribution shown in Fig.~\ref{dust_distribution} and SED fitting (Fig.~\ref{SEDs}), the best models to reproduce the SR~21 features are $v_f=30\rm{m~s}^{-1}$ and $\alpha_{\rm{turb}}=10^{-2}$ for outer planet masses of $M_{P2}=\{10,15\}~M_{\rm{Jup}}$. Figure~\ref{images_10Mjup} shows the HiCIAO and ALMA observations for one of these cases: the outer planet of mass $M_{P2}=10~M_{\rm{Jup}}$ where we see that the distribution of small particles is uniform, while the ALMA observations only show the ring caused by the outer dust trap, in reasonable good agreement with observations of SR~21. 

In the synthetic images of Fig.~\ref{images_10Mjup}, we do not reproduce the potential asymmetry due to eccentricity, since we only conduct dust evolution models in the radial direction by azimuthally averaging the results from hydrodynamical simulations.  As explained in the previous section, the asymmetry in flux between the upper and lower regions of the disk in the HiCIAO images is caused by the strong forward scattering of the small dust particles due to the inclination of the disk ($i=15^{\circ}$). Although low inclination, it is enough to make this effect noticeable. This is an effect not detected by the \citet{follette2013} observations which can be attributed to the fact that, although we have used realistic composition and size distribution for the dust particles, these are in our simulations smooth particles whereas real dust particles are expected to have rough surfaces.  Particles with rough surfaces have a flatter phase function at high viewing angles ($90\pm15$ in this case) and as a consequence the forward scattering of real particles is expected to be lower than that of the smooth particles used for these simulations \citep{minprep}. This would cause the asymmetry in flux to be reduced in real observations.

\section{Summary}     \label{summary}

Transition disks reveal intriguing structures such as spiral arms, asymmetric cavities, and spatial segregation between small and large grains. In this paper, by combining hydrodynamical and dust evolution models, we explore the gas and dust density distributions when two planets are embedded in a disk. After computing the radiative transfer, we predict the observational signatures in disks with various disk viscosity ($\alpha_{\rm{turb}}$), fragmentation velocity threshold ($v_f$), and planet mass. In particular, we investigate the possibility of having a smooth distribution of micron size particles over the entire disk and a large cavity in millimetre grains, as recently observed in transition disks, such as e.g. SR~21 \citep{follette2013, perez2014}, HD~169142 \citep{osorio2014}, and MWC~758. \citep{andrews2011, grady2013}. We also look for possible asymmetries in the gas distribution. 

Our framework is the following:  we fix a $1~M_{\rm{Jup}}$ planet in the inner disk at $r_{P1}$  around  a  $2.5~M_{\odot}$ star and a more massive planet further out in the disk at $3.5\times r_{P1}$, $\alpha_{\rm{turb}}=[10^{-3}, 10^{-2}]$, $v_f=[10,30]\rm{m~s}^{-1}$, and $M_{P2}=[5,10,15]~M_{\rm{Jup}}$. Our conclusions are the following (see also Tables \ref{results1} and \ref{results2}): 

\begin{itemize}
\item We confirm that for specific values of $\alpha_{\rm{turb}}$ and $v_f$,  trapping at pressure maxima is possible as a result of planets embedded in the disk \citep{pinilla2012}. Although trapping is very efficient for $\alpha_{\rm{turb}}\sim10^{-3}$, for higher turbulence ($\alpha_{\rm{turb}}=10^{-2}$), dust accumulation only occurs for massive planets ($M_{P2}=[10,15]M_{\rm{Jup}}$) and low fragmentation efficiency (i.e. high threshold velocities for destructive collisions). The sharpness of the ring-like structure seen at millimetre wavelengths strongly depends on $\alpha_{\rm{turb}}$ (Fig.~\ref{dust_distribution}). 

\item When trapping is efficient, a smooth distribution of micron-sized particles over the entire disk can only be observed if the combination of planet mass and turbulence is such that small grains are not fully filtered out. For $\alpha_{\rm{turb}}\sim10^{-3}$, an outer planet more massive than $5~M_{\rm{Jup}}$ will lead to micron-grain filtration. On the other hand, with $\alpha_{\rm{turb}}\sim10^{-2}$,  the disk is very diffusive, and micron-sized particles are distributed throughout the disk.

\item Depending on the disk viscosity and planet masses, asymmetries in the gas may exist at the outer edges of the gaps due to disk eccentricity or vortex formation. Eccentricities are induced by massive planets \citep{kley2006}, while long-lived vortices are created by massive planets in a low viscosity disk (Fig.~\ref{vor_ecc}).  The simulations presented in this paper show that the cases in which particle trapping occurs \textit{and} small particles are spread over the entire disk are also likely to show gas asymmetries caused by  disk eccentricity. In this case, it is expected that the azimuthal dust distribution of all particles is identical to gas distribution.

\item In the particular case of SR~21 transition disk, the observed cavity at millimetre wavelength \citep{andrews2011} can result from a planet located around 17-18AU. In this case an inner planet of $1M_{\rm {Jup}}$  at 5~AU can create a ring of gas around 6-7AU, which is in agreement with CRIRES observations of the gas \citep{pontoppidan2008},  where dust particles can also accumulate and lead to an NIR excess. To obtain micron-sized particles in the entire disk, the outer planet should not be too massive ($M_{P2}\approx5M_{\rm{Jup}}$), and $\alpha_{\rm{turb}}=10^{-3}$, $v_f=10~\rm{m}~s^{-1}$. If the disk viscosity is higher, the mass of the outer planet must increase to have effective trapping  ($M_{P2}=[10,15]M_{\rm{Jup}}$), and therefore a ring-like emission at mm-wavelengths.  In such cases, the fragmentation velocities should be velocities need to be significantly higher ($v_f=30~\rm{m}~s^{-1}$ than for bare silicate grains (few $\rm{m}~s^{-1}$).  These fragmentation velocities are expected for icy grains \citep{blum2008, wada2009}. For any of these cases, the resulting asymmetry is likely caused by eccentricity and it has low contrast of around $\sim 1.5-2$ in the gas, which may be enhanced  by spiral arms.

\end{itemize}

\begin{acknowledgements}
The authors are thankful to H.~Meheut for fruitful discussions and D.~Fedele for providing the DIGIT data of SR~21.  P.~P. is supported by Koninklijke Nederlandse Akademie van Wetenschappen (KNAW) professor prize to Ewine van Dishoeck. T.~B. acknowledges support from NASA Origins of Solar Systems grant NNX12AJ04G. Astrochemistry in Leiden is supported by the Netherlands Research
School for Astronomy (NOVA).
\end{acknowledgements}

\bibliographystyle{aa}

\bibliography{pinilla_2planets}

\end{document}